\documentclass{emulateapj}

\shorttitle{Eclipsing binaries in the MACHO database}
\shortauthors{Derekas et al.}

\begin{document}

\title{Ellipsoidal Variability and Long Secondary Periods in MACHO Red Giant
Stars}

\author{A. Derekas\altaffilmark{1}, L. L. Kiss\altaffilmark{1}, 
T. R. Bedding\altaffilmark{1}, H. Kjeldsen\altaffilmark{2}, 
P. Lah\altaffilmark{3},  Gy. M. 
Szab\'o\altaffilmark{4,5}}
\altaffiltext{1}{School of Physics, University of Sydney, NSW 2006, Australia; 
E-mail: derekas@physics.usyd.edu.au}
\altaffiltext{2}{Department of Physics and Astronomy, University of Aarhus,
DK-8000 Aarhus C, Denmark} 
\altaffiltext{3}{Research School of Astronomy \& Astrophysics, Australian 
National University, Canberra, Australia}
\altaffiltext{4}{Department of Experimental Physics, University of Szeged, 
D\'om t\'er 9. Szeged 6720, Hungary}
\altaffiltext{5}{Magyary Zolt\'an Postdoctoral Research Fellow}

\begin{abstract}

We present a period-luminosity-amplitude analysis of 5899 red giant and binary
stars in the Large Magellanic Cloud, using publicly available observations of
the MACHO project. For each star, we determined new periods, which were
double-checked in order to exclude aliases and false periods. The
period-luminosity relations confirm the existence of a short-period,
small-amplitude P--L sequence at periods shortward of Seq. A. We point out 
that the widely accepted sequence of eclipsing binaries between  Seqs. C and D,
known as Seq. E, does not exist. The correct position for  Seq. E is at periods
a factor of two greater, and the few stars genuinely lying  between Seq. C and
D are under-luminous Mira variables, presumably  enshrouded in dust. The true
Seq. E overlaps with the sequence of Long Secondary Periods (Seq. D) and their
P--L relation is well described by a simple model assuming Roche geometry. The
amplitudes of LSPs have properties that are different from both the pulsations
and the ellipsoidal variations, but they are more similar to the former than
the latter, arguing for pulsation rather than binarity as the origin of the LSP
phenomenon.

\end{abstract}

\keywords{binaries: eclipsing --- galaxies: individual (Large Magellanic Cloud)
--- stars: AGB and post-AGB --- stars: oscillations --- stars: statistics --- 
stars: variables: other}

\section{Introduction}

The multiplicity of red giant period-luminosity (P--L) relations has been a
major discovery on the road to interpreting complex light variations of these
stars. Following the two seminal papers by \citet{woo99} and \citet{woo00}, a
picture has emerged that can be summarized as follows: large-amplitude Mira
stars pulsate in the fundamental mode, whereas smaller-amplitude semiregulars
are often multimode pulsators, in which various overtone modes can be excited
\citep[see also][]{bed98}. Besides the pulsating P--L sequences (Seq. A, B and
C, as labeled by \citet{woo99}), two other sequences were suggested: Seq. E
with red giants in eclipsing binaries and Seq. D with stars that have long
secondary periods (LSPs). The latter pose a great mystery and the nature of
their slow variations is still not understood, with several different
mechanisms proposed \citep{oli03,woo04}.

The basic picture of multiple P--L relations has been confirmed by many
independent studies, mostly based on $K$-band magnitudes. It has emerged that
the original five sequences have further details, including a break at the tip
of the Red Giant Branch (RGB), which is due to the existence of distinct RGB
pulsators that are mixed with the more evolved AGB variables
\citep[e.g.][]{ita02,kis03,kis04,ita04a,ita04b,sos04a,fra05}. Almost all
authors have accepted the existence of the distinct sequences of red giant
binaries (Seq. E) and LSP stars (Seq. D). The only exception was
\citet{sos04b}, who showed that Seqs. E and D seem to merge at a specific
luminosity (as measured by the Wesenheit index) and suggested that this may
imply the binary origin of LSPs.

Here we report on a combined analysis of MACHO observations of eclipsing
binaries and red giants in the Large Magellanic Cloud, which shed new light on
these stars and on the LSP phenomenon.

\section{Data analysis and results}

Our results are based on two sets of publicly available almost eight-year long
MACHO light curves. A detailed description of the MACHO project can be found in
\citet{coo95}. Some of the data are offered for download through the MACHO
website (http://wwwmacho.mcmaster.ca), where one can choose specific samples
based on an automated classification of variability type. Using the Web
interface, we individually downloaded all light curves classified as eclipsing
binaries (6833 stars) and as red giant variables (classified as Wood A, B, C
and D classes; 2868 stars).

In the case of the eclipsing binaries, it became obvious very quickly that the
classification was not perfect, and a large fraction of stars turned out to be
Cepheids, RR Lyrae stars or long-period variables. We also found that the
catalogued periods were incorrect for a significant number of stars. We
therefore reclassified all 6833 stars and re-determined their periods, using
the following procedure (more details will be given elsewhere). 

Periods were first estimated using the Phase Dispersion Minimization method
\citep{ste78}. We then checked all the folded light curves by eye and refined
the periods with the String--Length method \citep{laf65,cla02}, which is more
reliable than PDM when the light curve contains long flat sections and very
narrow minima, as is the case for many eclipsing binaries. Also, in many cases
PDM gave harmonics or subharmonics of the true period, which was only
recognized through the visual inspection of every phase diagram. We also
examined the color variations to identify and exclude pulsating stars with
sinusoidal light curves. After this analysis, 3031 stars remained as genuine
eclipsing or ellipsoidal variables. 

\begin{figure*}  
\epsscale{1}
\plottwo{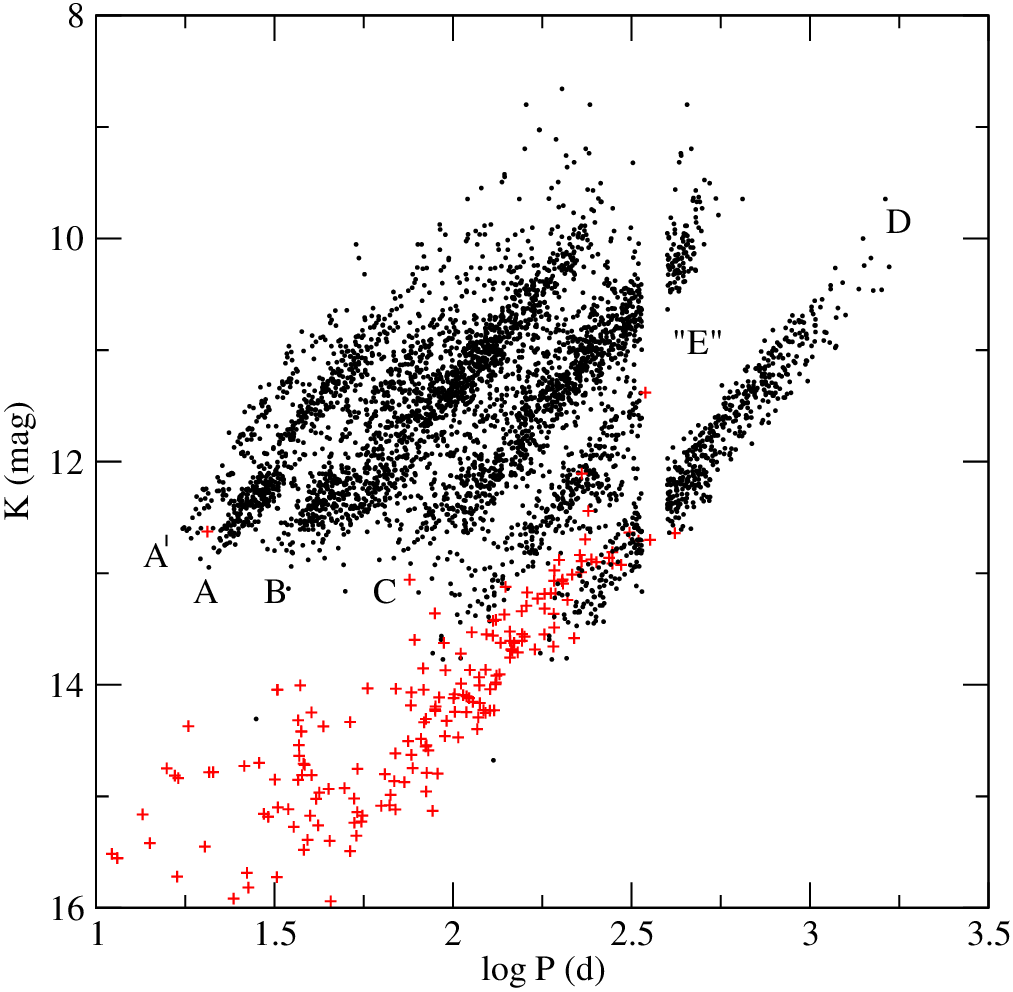}{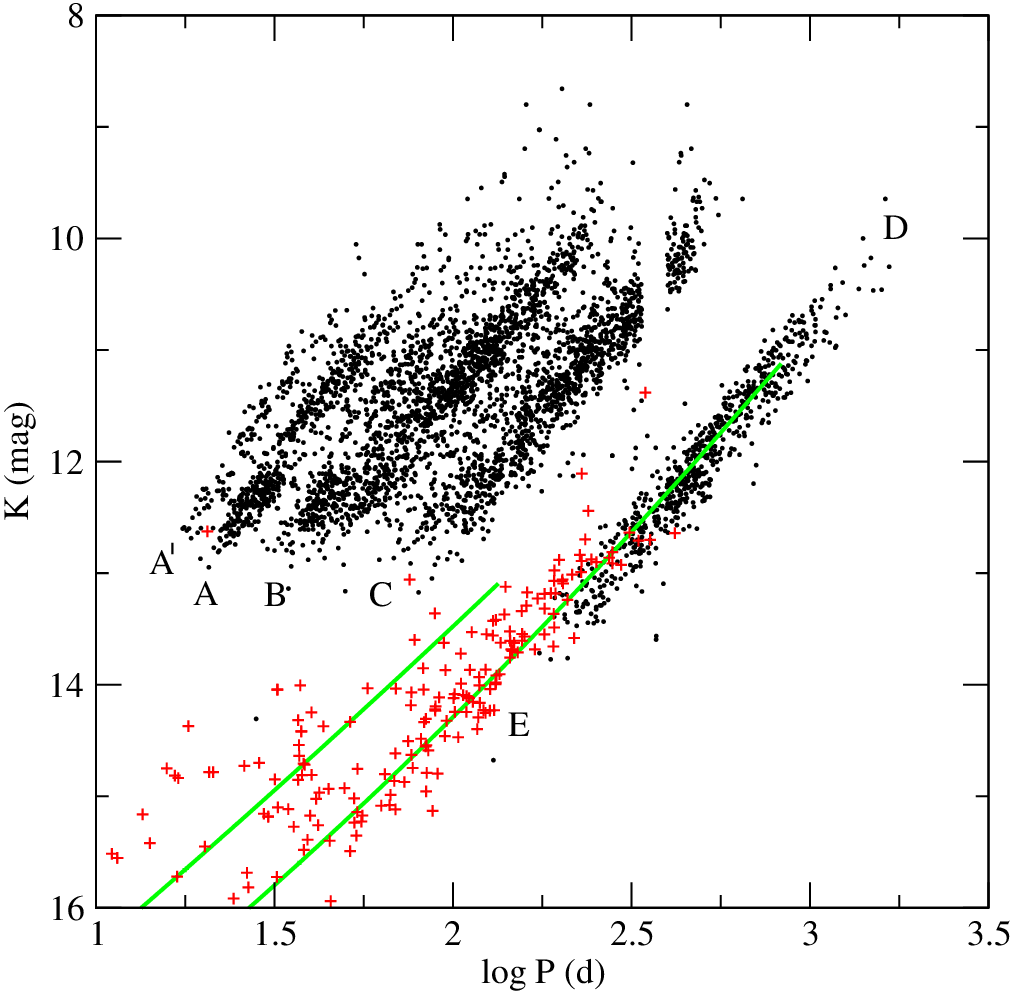}
\caption[]{\label{pl}Period-Luminosity relations for red giant pulsators (black
dots) before (left panel) and after (right panel) period correction of stars on 
Seq. E; 310 ellipsoidal and contact eclipsing binaries (red pluses) 
are shown with the correct periods in both plots.
The solid lines in the right panel show model calculations using evolutionary 
tracks and assuming Roche-geometry (see text).} 
\end{figure*}

Next, we classified the binary sample using Fourier decomposition of their
phase diagrams. \citet{ruc93} showed that light curves of W UMa systems
(contact binaries) can be quantitatively described using only two coefficients,
${\rm a_{2}}$ and ${\rm a_{4}}$, of the cosine decomposition ${\rm \sum a_{i}
\cos (2\pi i \varphi)}$. \citet{poj02} tested the behavior of semi-detached and
detached systems in the ${\rm a_{2} - a_{4}}$ plane by decomposing theoretical
light curves into Fourier coefficients. We found that only stars with  ``W
UMa-like'' light curve shape composed the sequence (which is plotted in Fig.\
\ref{pl}), while detached and semi-detached systems are spread everywhere in
the P--L plane.

Our second set of light curves were those of the 2868 publicly available MACHO
red giant variables. Since they often show multiply periodic light variations,
we determined periods with iterative sine wave fitting. As a measure of
significance, we also estimated the $S/N$ ratio of the peaks in the Fourier
spectra. Since the noise in the Fourier spectrum increased toward lower
frequencies, different values of the $S/N$ were used when determining whether a
peak was real for different period-luminosity relations ($S/N$ cutoff was set
to 3 for Seq. A$^\prime$, 4 for Seq. A, 6 for Seqs. B, C and 10 for Seq. D). We
omitted periods close to 1 yr, because many light curves show variations with
this period that are not real. As a result, a total of 4315 significant
frequencies were identified for the 2868 stars.

We studied these two samples in the P--L plane. In order to reduce the effects
of interstellar extinction and allow a direct comparison with previous results
in the literature, we plotted the period--$K$ magnitude relation. We obtained
near infrared magnitudes by cross-correlation with the 2MASS All-Sky Point
Source Catalog (http://irsa.ipac.caltech.edu), with a search radius of
3$^{\prime\prime}$. The resulting P--L diagram is shown in the left panel of
Fig.\ \ref{pl}. 

To our surprise, the P--L relation of the binary sample did not follow Seq. E,
as we had expected. Instead, they overlapped with Seq. D, which at first sight
appears to give strong evidence for the binary origin of Seq. D and prompted us
to investigate the issue in more detail.

\section{Discussion}

The combined P--L plot in Fig.\ \ref{pl} shows the well-known complex structure
of distinct sequences. Besides sequences A, B, C, D and E, we also detect the
existence of the faintly visible new short-period P--L sequence (Seq.
A$^\prime$) on the left-hand side boundary of the diagram (labeled as a$_{4}$
and b$_{4}$ in \citet{sos04a} and $P_4$ in \citet{kis06}). The eclipsing stars
(pluses) seem to merge with Seq. D rather than forming Seq. E as adopted in the
literature \citep{woo99,woo00,kis03,kis04,ita04a,ita04b,nod04,fra05}. 

\begin{figure}   
\epsscale{0.9} 
\plotone{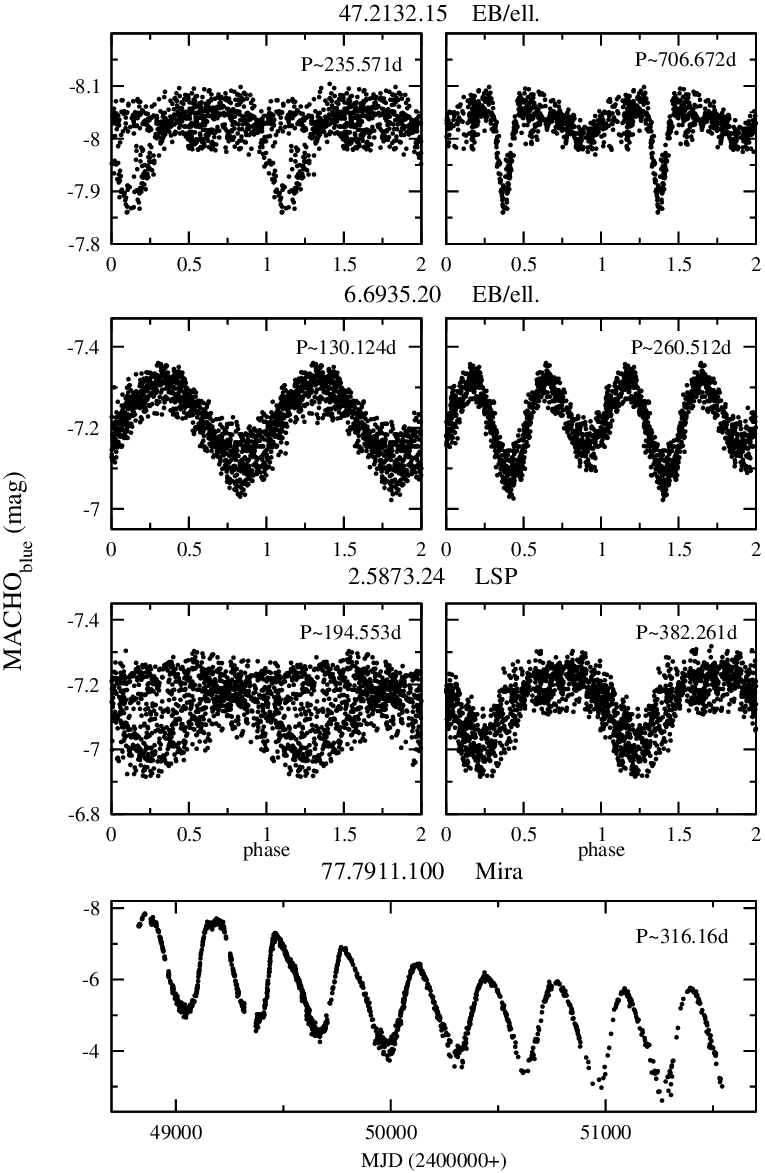} 
\caption[]{\label{example}
Example of phase diagrams before (left panels) and  after (right panel) period
correction. The first and the second star from the top show eclipsing binary
and ellipsoidal variable light curve, while the third  star has a typical LSP
variation. The bottom panel shows the light curve of a Mira star that lies in
the gap between Seqs. C and D and which faded $\sim$2 mags during 8 years of
MACHO observations.}  
\end{figure}

To clarify this issue, we re-checked periods for: (i) stars on Seq. E in Fig. 1
of \citet{woo00}, for which the identifiers and basic data were kindly provided
by Peter Wood; (ii) stars on Seq. E in our Fig.\ \ref{pl} (black dots). In both
cases, it turned out that for most of the objects, the given periods were half
of the true ones, as one might expect from a Fourier analysis of eclipsing
binary light curves. We have carefully double-checked all individual light
curves on Seq. E and D and corrected the periods (see Fig.\ \ref{example}). The
final P--L plot is shown in the right panel of Fig.\ \ref{pl}.

As a result of the period correction, the sequence between C and D, which is
known as Seq. E in the literature, has completely disappeared. A few stars
remain in the gap but practically none of them are eclipsing binaries. The
majority turn out to be Mira stars that are very red ($J-K>2$ mag) and are
presumably carbon-rich Miras that are dimmed by circumstellar dust clouds
(bottom panel of Fig.\ \ref{example}). We propose to retain the label E for the
sequence of ellipsoidal variables in its corrected position (right panel of
Fig.\ \ref{pl}).

In the literature, only \citet{sos04b} plotted the eclipsing binary sequence at
the correct (doubled) period. However, they did not discuss this issue  and one
can still find more recent studies where Seq. E  was shown at the wrong period.
\citet{sos04b} have, however, shown that Roche geometry gives a good fit to the
OGLE ellipsoidal variables. We have also checked this on the MACHO sample. We
calculated the theoretical orbital periods of systems at mass ratios of 1,
where the components fill their Roche lobes. For this, we used evolutionary
models of \citet{cas03} and applied equations of Section 4 in \citet{sos04b}.
(Note, however, that the definition of $f(q)$ in their Eq. 1 actually gave
$f(q)^{-1}$ as the filling factor.) For the calculations we took the
evolutionary tracks of the 0.85 ${\rm M}_{\odot}$ and 2.5 ${\rm M}_{\odot}$
models, since theses masses represent the mass limits of stars that evolve
through the RGB and AGB. The $K$ magnitudes of the models were determined from
$T_{\rm eff}$ vs. $(V-K)$ calibrations, combined from \citet{hou00} and
\citet{kuc06}. 

In the right panel of Fig.\ \ref{pl} we show these two limits (the shorter
period line belongs to the higher mass). For any other mass ratios, the orbital
periods shift towards smaller values. This simple approach with Roche geometry
describes remarkably well the observed period-luminosity relation of
ellipsoidal variables in the MACHO sample, in agreement with the study of
\citet{sos04b}. However, the OGLE sample contains a larger fraction of
ellipsoidals and most of them have longer periods than are predicted by the
models. Those stars do not entirely fill the Roche lobe \citep{sos04b}. It is
also worth mentioning that the low fraction of ellipsoidal/eclipsing RGB stars
in the MACHO data (1.5\%) was used by \citet[]{woo04} as an argument against
the binary origin of LSPs. However, the OGLE statistics clearly showed that
there are at least 10 times more such RGB stars, and so that argument is no
longer valid.

Since the lower mass model fits Seq. D quite well, the question arises: how
similar are the ellipsoidal and LSP variables? To assess this, we examined the
amplitudes of these stars. \citet{sos04b} mentioned that amplitudes of LSPs are
positively correlated with the brightness of the star. Compared to OGLE data,
the MACHO observations have the advantage of giving information on the color
variations. We examined the amplitudes in the MACHO blue and red bands of the
binary and LSP stars, and also included Seq. C stars, which allow a comparison
with stars that we know to be pulsating.

\begin{figure}    
\epsscale{0.9}  
\plotone{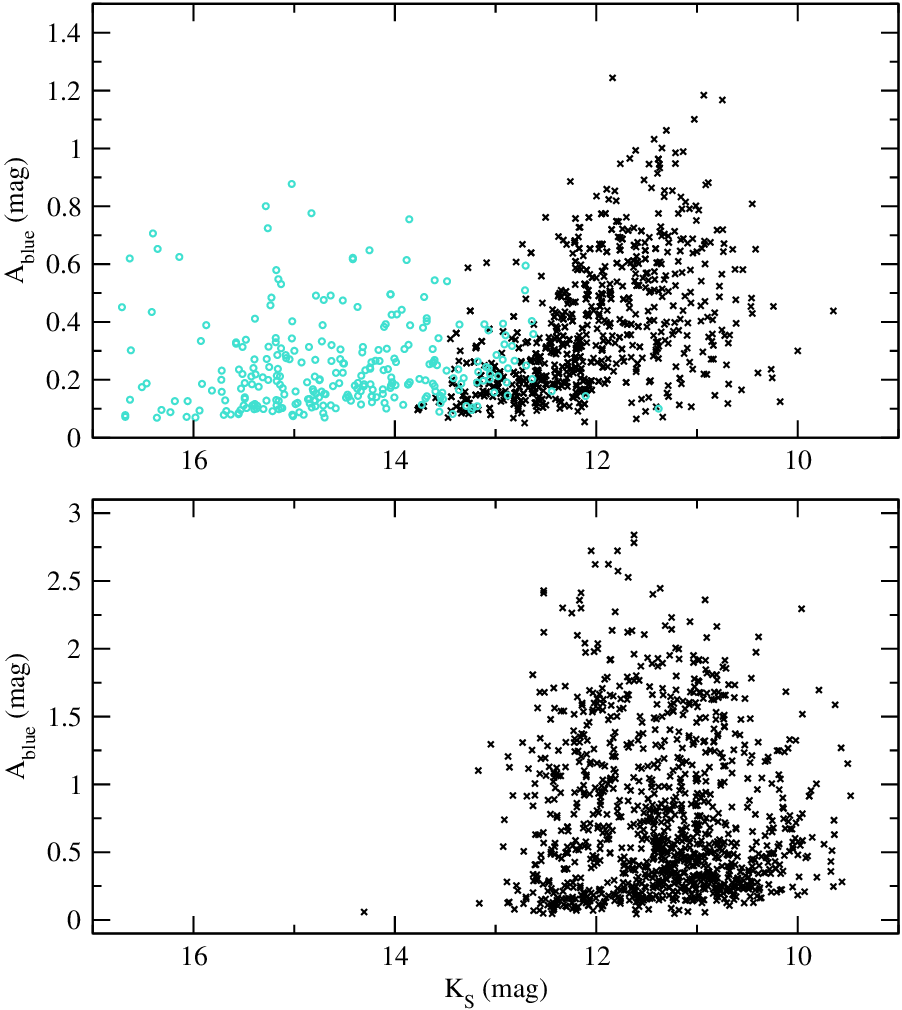} 
\caption[]{\label{amlu} Luminosity-amplitude relation of stars on Seq. C, D 
and E. {\it Top panel:} Black crosses show the luminosity-amplitude relation 
for Seq. D stars, while turquoise circles the same for Seq. E stars, which 
contain  ellipsoidal variables. The difference of the two sequence is very 
clear, e.g. amplitudes of Seq. D stars show moderately positive correlation 
with the  luminosity. {\it Bottom panel:}  Luminosity-amplitude relation for 
Seq. C stars, which are fundamental mode pulsators. This relation is
significantly different from those of  Seqs. D and E stars.}   
\end{figure}

The amplitudes were measured by fitting smooth spline functions to the phased
light curves. The resulting blue peak-to-peak amplitudes are shown as a
function of $K$ magnitude in Fig.\ \ref{amlu}, where several features are
apparent. Firstly, the amplitudes of ellipsoidal variables (circles in the
upper panel) do not show any correlation with luminosity, as expected for
variation caused by the geometry of a binary system. Secondly, the amplitudes
of LSP variables (crosses in the upper panel) increase with luminosity, with
their distribution forming a striking triangular envelope (the correlation
coefficient is $r\sim0.51$). It is also very interesting that the points below
the well-defined upper envelope seem to show a flat distribution. Whatever the
cause of the LSP phenomenon, there appears to be a maximum possible  amplitude
at each luminosity, with an apparently uniform distribution of  amplitudes
beneath this maximum value.

In comparison, stars on the pulsating P--L sequence C (lower panel in Fig.\
\ref{amlu}), show a very different distribution. To quantify the difference
between the amplitude-luminosity distribution of Seqs. C and D, we performed
the multivariate $\varepsilon$-test of \citet{sze04}, applied for two
dimensions. In this test we measure the effects of random permutations of the
initial distributions via changes in the ``information energy'' of the two
distributions. In our case, 1000 random permutations showed that the difference
between the C and D samples is highly significant. Therefore, we conclude that
the physical mechanism causing the LSP phenomenon must be different from both
the ellipsoidal variations of Seq. E and the radial fundamental-mode 
pulsations of Seq. C. Note, however, that for Seqs. A and B, there is a similar
positive correlation between amplitude and luminosity (see the upper three
panels in figs. 4 in \citet{kis03} and \citet{kis04}) than for Seq. D. 

\begin{figure}   
\epsscale{0.9} 
\plotone{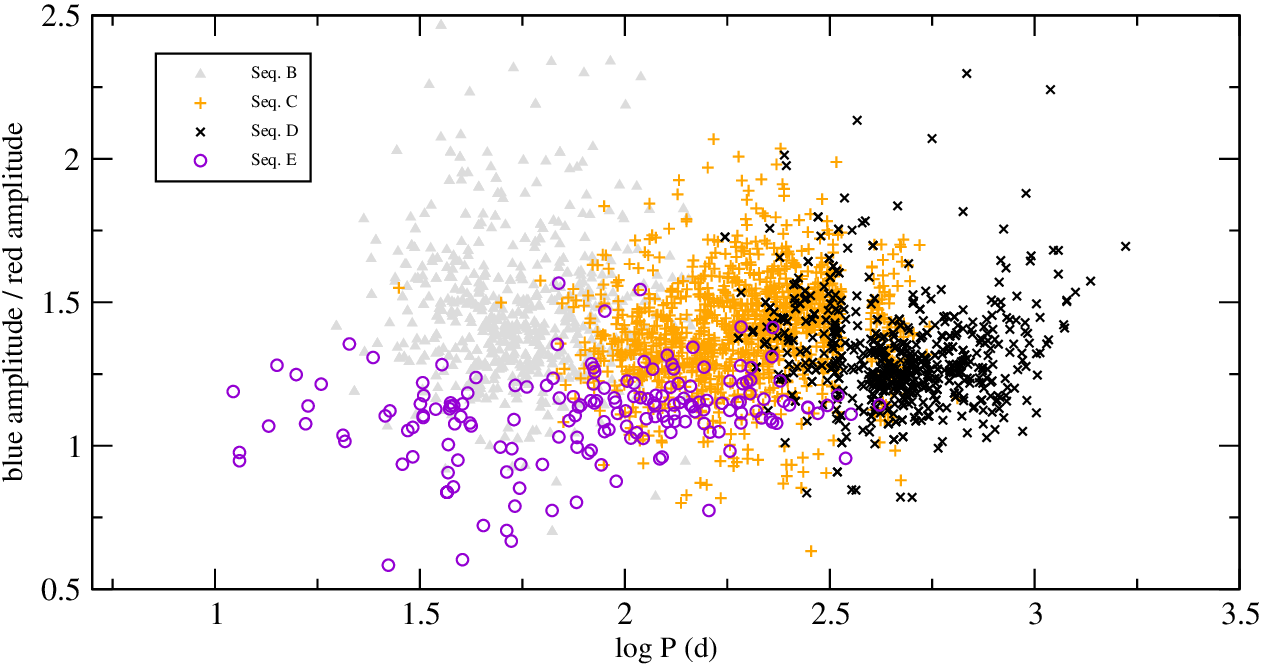} 
\caption[]{\label{pamp} Blue-to-red amplitude ratios as function of period for 
stars on Seqs. B (grey triangles), C (orange pluses), D (black crosses) and E
(purple circles). Grey  triangles show those short period pulsations of Seq. D
stars, which fall to  Seq. B. The two clearly pulsating sequences, Seqs. B and
C are the continuation  of each other. Ellipsoidal variables (Seq. E) have
lower amplitude ratio than  pulsating stars. Amplitude ratios of Seq. D stars
are more similar to those of pulsating stars.} 
\end{figure}

At the same time, comparing blue and red amplitudes for individual stars
revealed further interesting information. In Fig.\ \ref{pamp} we plot the
blue-to-red amplitude ratios as function of period for Seqs. B, C, D and E
stars. For the pulsating objects the median of the ratio is 1.40, indicating
strong color, thus temperature, changes during the pulsations. For
ellipsoidal/eclipsing binaries, the median ratio is 1.13, while the LSPs have a
median  ratio of 1.29, being more similar to the pulsating stars \citep[see
also][]{hub03}. This behavior agrees with the findings of \citet{woo04}
who, based on color-amplitude variations of single objects, argued for the
pulsational origin of LSPs. The overall statistics of more than 700 LSP
variables favors this argument over the binary hypothesis, also agreeing with
\citet{hin02}, who concluded that the long-period velocity changes in their
observed stars probably result from some kind of pulsation.

\section{Summary}

The main results of this paper can be summarized as follows:

\begin{list}{$\cdot$}{}
 {\parskip=-6pt

\item the period-luminosity relations of $\sim$6000 stars based
on MACHO data confirm the existence of the short-period, small-amplitude P--L
sequence at shortward of Seq. A, which belongs to a higher-overtone pulsation
mode. We label this Seq. A$^{\prime}$.

\item the widely accepted sequence of eclipsing binaries between C and D, known as
Seq. E, does not exist. The correct position for Seq. E, which comprises contact
binaries and ellipsoidal variables, is at periods a factor of two greater. The true
Seq. E overlaps with the LSPs (Seq. D), which appears to suggest a binary origin
for the LSP phenomenon (but see the last point).

\item of the few stars that genuinely lie between Seq. C and D, most are
under-luminous Mira variables, presumably enshrouded in dust.

\item we confirmed that ellipsoidal variables have a similar P--L relation to
LSP stars. Their P--L relation is well described by a simple model assuming
Roche geometry.

\item the amplitudes of LSPs have properties that are different from both the
pulsations and the ellipsoidal variations, but they are more similar to the
former than the latter, arguing for pulsation rather than binarity as the
origin of the LSP phenomenon.

 }
\end{list}

\acknowledgments

AD is supported by an Australian Postgraduate Research Award. LLK is supported by a
University of Sydney Postdoctoral Research Fellowship. GyMSz was supported by the
Magyary Zolt\'an Higher Educational Public Foundation. This paper utilizes public
domain data obtained by the MACHO Project, jointly funded by the US Department of
Energy through the University of California, Lawrence Livermore National Laboratory
under contract No. W-7405-Eng-48, by the National Science Foundation through the
Center for Particle Astrophysics of the University of California under cooperative
agreement AST-8809616, and by the Mount Stromlo and Siding Spring Observatory, part
of the Australian National University.

\clearpage

\end{document}